\documentclass[12pt]{article}

 \usepackage{amsfonts}
 \usepackage{amssymb,amsmath}
		 \usepackage{graphicx} \usepackage{epstopdf}  
 \numberwithin{equation}{section}

 \newcommand{\crlb}[1]{\label{#1}\\[2pt]}
 
 \newcommand{\crld}[1]{\label{#1}}
 \newcommand{\eela}[1]{\quad\hbox{\scriptsize{#1}}\label{#1}\end{eqnarray}}
 \newcommand{\eelb}[1]{\label{#1}\end{eqnarray}}
 
 \newcommand{\newsecb}[2]{\section{#1}\label{#2}\setcounter{equation}{0}}

 \newcommand{\nolabels} {\def\eel{\eelb}\def\eeql{\eeqlb}  \def\crl{\crlb} \def\newsecl{\newsecb}\def\bibiteml{\bibitem} \def\citel{\cite}\def\labell{\crld}}
\newcommand{\eeqla}[1]{\quad\hbox{\scriptsize{#1}}\label{#1}\end{aligned}\end{equation}}
\newcommand{\eeqlb}[1]{\label{#1}\end{aligned}\end{equation}}

\newcommand\publishversion{\nolabels\setlength{\textheight}{8.3in}\setlength{\oddsidemargin}{0in}
   	 \setlength{\textwidth}{6.3in}\setlength{\topmargin}{-0.2in}}

\def\beq{\begin{equation}\begin{aligned}}		\def\eeq{\end{aligned}\end{equation}}
\def\be{\begin{eqnarray}}  					\def\ee{\end{eqnarray}}		
\def\bi#1{\begin{itemize}\item[#1]} 			\def\itm#1{\item[#1]} 			\def\ei{\end{itemize}}
  \def\eqn#1{(\ref{#1})}
   	 \def\fn{\footnote}	  		\def\nm{\nonumber}

		 \def\del{\delta}  

 \def\del{\delta}         
             
             \def\vv{\varphi}    
             \def\rr{\varrho}

 \def\w{\omega}

 \def\pa{\partial} \def\ra{\rightarrow} 
  
 \def\dd{{\rm d}}  \def\bra{\langle}   \def\ket{\rangle}
\def\qu{\overset{\textstyle ?}{=}}

\def\fract#1#2{{\textstyle\frac{#1}{#2}}}	 	 	
\def\ffract#1#2{\raise .2 em\hbox{$\scriptstyle#1\,$}\kern-.34 em/\kern-.34 em\lower .15 em \hbox{$\scriptstyle\,#2$}}
\def\half{\fract12}		\def\quart{\fract14}			
\def\tl#1{\tilde{#1}} 
\def\ex#1{e^{\textstyle#1}} 		\def\qqquad{\qquad\qquad}	
\def\bpmatriÄx{\begin{pmatrix}} 			\def\epmatrix{\end{pmatrix}}
\def\bmatrix{\begin{matrix}} 			\def\ematrix{\end{matrix}} 
\def\bcenter{\begin{center}}			\def\ecenter{\end{center}}

\def\lowerheightgth#1#2#3{\(\raise-#1\hbox{\includegraphics[height=#2]{#3}}\)}
\def\lowerwidthgth#1#2#3{\(\raise-#1\hbox{\includegraphics[width=#2]{#3}}\)}
\def\widthfig#1#2{\includegraphics[width=#1]{#2}}

\def\ontt{{\mathrm{ont}}}

\def\weglaten#1{}	
 
   \def\ret{\\[5pt]} 
\def\adag{{a^\dag}} 

\publishversion 
 
\begin{document}
\begin{titlepage}
 \title{The Hidden Ontological Variable in \\ Quantum Harmonic Oscillators\ret }
 \author{Gerard 't~Hooft}
\date{\normalsize Institute for Theoretical Physics \\ Utrecht University  \\[10pt]
Princetonplein 5 \\
3584 CC Utrecht \\
 the Netherlands  } \maketitle 
 
{ \abstract{\begin{quotation}The standard quantum mechanical harmonic oscillator has an exact, dual relationship with a completely classical system: a classical particle running along a circle. Duality here means that there is a one-to-one relation between all observables in one model, and the observables of the other model. Thus the duality we find, appears to be in conflict with the usual assertion that classical theories can never reproduce quantum effects as observed in many quantum models. 
We suggest that there must be more of such relationships, but we study only this one as a prototype. It reveals how classical ``hidden variables'' may work. The classical states can form the basis of Hilbert space that can be adopted in describing the quantum model. Wave functions in the quantum system generate probability distributions in the classical one. One finds that, where the classical system always obeys the rule ``probability in = probability out'', the same probabilities are quantum probabilities in the quantum system. It is shown how the quantum x and p operators in a quantum oscillator can be given a classical meaning. It is explained how an apparent clash with quantum logic can be explained away.
\end{quotation}}}
 \end{titlepage} 
\vfil

\newsecl{Introduction}{intro.sec}
It has become customary to investigate quantum theories by proving that they cannot be represented in terms of ontological variables. These ontological variables, known as `local hidden variables' (LHV), are assumed to reproduce the results of all experiments that can be performed on a given quantum system, which is subsequently shown to lead to logical contradictions.

However, when the outcome of an extensively examined quantum experiment is compared with a classical theory, it is often the classical dynamics that is finished off in one short sentence: ``This cannot be the result of  a classical theory''. 
One may however suspect that the assumptions made concerning these LHV are too strict, so that there could be loopholes\fn{A very important loophole, not discussed further in this paper, is that models such as the Standard Model of the elementary particles, require perturbation expansions, which are known  to be fundamentally divergent. This procedure introduces uncertainties\cite{GtHFF.ref} that can be studied further, under the suspicion that this could be the cause of the tendency of quantum wave functions to spread.}. Many investigations are aimed at closing  these loopholes by making further assumptions.\cite{Bell.ref}--\cite{GHZ.ref}

This, we claim, may not be the only way to improve our understanding of quantum mechanics. Here, we approach the question concerning the interpretation of quantum mechanics from the other end: which quantum systems  \emph{do} allow for classical variables, and can these models be extended to include physically useful ones? Can these models be demanded to obey (some form of) locality? Can we use them as building blocks? We claim that this is a rich field for further excursions\cite{Brans-1987.ref}--\cite{GtHmassive.ref}.  Here,  a very important example is  exhibited: the quantum harmonic oscillator. As we shall see, it contains a variable that can explain everything we see in a quantum harmonic oscillator, in terms of completely classical mathematical logic. Our variables are not hidden at all, and completely ontological; therefore we call our variable `COV', standing for ``Classical Ontological Variable''. The letter \(L\) is omitted, since locality may not be guaranteed, and anyway, we do not intend to contradict earlier no-go theorems, but rather search for ways out\fn{Locality is not a meaningful concept for the single harmonic oscillator.}. Understanding the COV may be an important alley that could lead us to new insights, perhaps even in model building.\cite{Jegerlehner.ref},\cite{GtHYang.ref}.\setcounter{page}2

The most important part of this paper is Section~\ref{harm.sec}. Here we show how any quantum harmonic oscillator, contains an ontological degree of freedom. Using modern jargon, we observe that the quantum harmonic oscillator is \emph{dual} to a classical particle on a circle. 

Questions asked after a talk presented at the Lindau Meeting, June/July 2024, made me realise that the features discussed below are not very well-known and therefore this short publication may be useful.

\newsecl{The harmonic oscillator}{harm.sec}

 In one space-like dimension, consider the Hamiltonian\fn{For convenience, we set the ground state energy to zero; ground-state energies can be returned whenever this might be needed.} \(H\) of an elementary quantum harmonic oscillator in terms of  the variables \(x\) and \(p\),
\be [x,p]=i\ ,\qquad H=\half(p^2+x^2-1)\ . \eel{harmops.eq}
Planck's constant will always be set as \(\hbar=1\), and as such it merely relates the units of energy  to the units of frequencies. Also the angular frequency \(\w\) is set to 1. The operator equations are
	\be \frac{\dd x}{\dd t}=i[H,x]=p\ ,\qquad \frac{\dd p}{\dd t}=i[H,p]=-x\ . \eel{queom.eq}

We  shall need the annihilation operator \(a\) and the creation operator \(\adag\), defined by
	\be a=\fract 1{\sqrt2}(x+ip)\ ,&& \adag=\fract 1{\sqrt2}(x-ip)\ ,\qquad [a,\adag]=1\ ,\nm \\
	x=\fract1{\sqrt2}(a+\adag)\ ,&& p=\fract i{\sqrt2}(\adag-a)\ ,\qquad [x,p]=i\ .
	\eel{createann.eq}
(For practical reasons, the signs chosen in our definitions, deviate from the signs chosen in other work).
The eigenstates \(|n\ket^E\) of \(H\), and their eigenvalues \(E_n\),  are found as usual to obey:
	\be H|n\ket^E=\adag a\,|n\ket^E = E_n|n\ket^E\ , \qquad E_n\ = \ n\ =\ 0,1, \cdots \eel{harmeigen.eq}

This, of course, is a completely standard, quantum mechanical procedure applied to the harmonic oscillator, but now we claim that it is \emph{dually related}  to a completely classical model. The classical system we have in mind is \emph{a particle moving on the unit circle}, with fixed velocity \(v=1\), and period \(=\ 2\pi\). The solution of its e.o.m. is: 
\be \vv(t)=\vv(0)+ t\mod 2\pi\  ;\eel{classmotion.eq}
 \(\vv\) is constrained to the interval \([0,\ 2\pi)\), where the boundary conditions are periodic.

To make our point, it is important to introduce (temporarily) a large integer \(N\), and a variable \(s=0,\cdots, \ N-1\,\), discretising the allowed values of \(\vv\), as follows:
\be \vv=2\pi s/N\ , \qquad s=0,\ 1,\ \cdots,\ N-1\  . \ee 
This matches with the introduction of small, finite time steps \(\del t=2\pi/N\,\). The  \(\vv\) states span an \(N\)-dimensional vector space \(\lbrace|s\ket^\ontt\rbrace\), where the superscript `ont' stands for ontological.

	 \begin{figure}[h!]   
	 
	\qqquad	\widthfig{270pt}{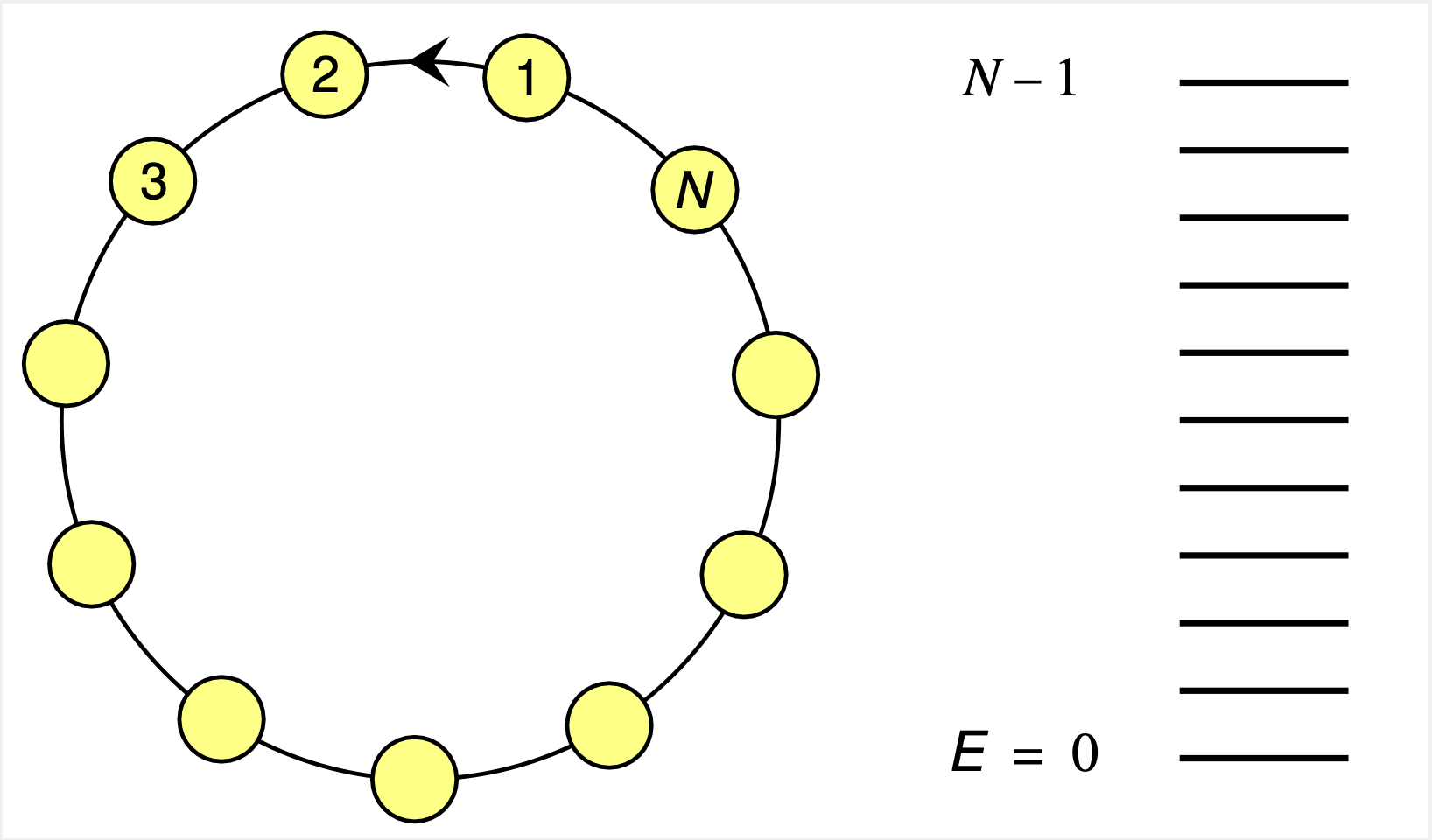}
		\caption{\qquad \small The ontological states \(|s\ket^\ontt\) when time is discrete,
			 \(\del t=\fract{2\pi}{N}\)\,. \newline\indent
			 {In this  picture, the choice  \(N=11\) was made. The energy spectrum  is shown; 
			the energies form the same sequence as in harmonic oscillators, in particular if we take 						the limit   \(N\ra\infty\).}} \labell{period.fig} 
	\end{figure}

	The energy eigenstates \(|n\ket^E\,,\ n=0\,,\ \cdots,\ N-1\,,\) of this rotating particle are superpositions of the ontological states:
\be |n\ket^E = \frac 1 {\sqrt {N}}\sum_{s=0}^{N-1}\ex{\frac{2\pi}{N}  i \,n\, s}\,|s \ket^\ontt\ . \eel{nE.eq}
 	 with the inverse:
\be |s\ket^\ontt = \frac 1 {\sqrt {N}}\sum_{n=0}^{N-1}\ex{-\frac{2\pi}{N}  i \,n\, s}\,|n\ket^E\ \eel{sont.eq}
	Note that these equations are merely discrete Fourier transformations.
	By checking the time dependence of \(|n\ket^E\) and \(|s\ket^\ontt\)\,, we see that
\be |n\ket^E\,(t)&=&\ex{-int}\,|n\ket^E\,(0)\labell{Etime.eq}\\  \hbox{and}\qquad |s \ket^\ontt\,(t)&=&|s-t/2\pi N\ket^\ontt\,(0)\ .\eel{stime.eq}
	We now note that the first \(N\) energy eigenstates of the harmonic oscillator, Eq.~\eqn{harmeigen.eq}, obey 
	exactly the same equations \eqn{Etime.eq}, and therefore Eqs.~\eqn{nE.eq} and \eqn{sont.eq} 
	define  \(N\) states, obeying \eqn{classmotion.eq}.

There is an important reason to start with a finite number \(N\). We see that, in these equations, the energy spectrum not only has a lowest energy state, \(|0\ket^E\), but also a highest energy state, \(|N-1\ket^E\). With strictly continuous angular variables \(|\vv\ket^\ontt\), we could postulate an energy spectrum running from \(-\infty\) to \(+\infty\). This would not dually correspond to a harmonic oscillator.\fn{At finite \(N\), there is an exact,  dual relationship to the \(SU(2)\) algebra, with \(N=2\ell+1\).} 

In this paper, we keep the lowest energy to be \(E=0\), while the highest energy will be unbounded. This enables us to take the limit \(N\ra\infty\), where we can write:
 \be \vv=2\pi s/N\,, \qquad\dd\vv=\frac{2\pi}N\ ,\qquad \hbox{and}\qquad  |\vv\ket^\ontt=|s\ket^\ontt/\sqrt{\dd\vv}\ ;\ee 
this turns Eqs~\eqn{nE.eq} and \eqn{sont.eq} into
\be\qquad |n\ket^E=\frac 1{\sqrt{2\pi}}\oint\dd\vv\,\ex{i\,\vv\,  n}\,|\vv\ket^\ontt\ , \qquad  |\vv \ket^\ontt=\frac 1{\sqrt{2\pi}}\sum_{n=0}^\infty \ex{-i\,\vv\,  n}\,|n\ket^E\ .\eel{contlim.eq}

Thus we proved that harmonic oscillators can be described in terms of  variables \(|\vv\ket^\ontt\) that evolve deterministically. It is easy to see that, due to Eq.~\eqn{classmotion.eq}, the wave function in terms of the \(s\) variable (or the \(\vv\) variable) does not spread.

However, the wave function may not have been chosen to collapse. In that case, the probability distribution \(\rr(\vv_1)=\big|\bra \vv_1|\vv\ket^\ontt\big|^2\) can be seen to rotate along the circle in the same way as \(\vv\) itself, so that we easily conclude that this probability distribution merely reflects the probabilities of the initial state.

This is a typical feature of the COV in a theory: these variables can be projected on the basis states of any Hilbert space, in which case the theory reproduces the probability distribution of the final states in terms of that of the initial
states. It is very important, however, that this identification between Hilbert space and the space of classical probability distributions, only applies to the \emph{ontological basis} of Hilbert space, that is, the basis spanned by all ontological states (the states \(|\vv\ket^\ontt\) in the case of the harmonic oscillator).
Thus we emphasise: any quantum harmonic oscillator is mathematically equivalent to a periodically  moving particle on a unit circle, and the wave function of a quantum harmonic oscillator merely reflects the probability distribution on this circle, if the initial state is not  known with infinite precision.
 
Some useful auxiliary functions are  \be G(z)\equiv\sum_{n=1}^\infty \sqrt n \,z^n\ ,\qquad\hbox{and}\qquad g(\vv)=G(\ex{i\vv})\ .\eel{auxfn.eq}
Since the annihilation operator \(a\), defined in Eq.~\eqn{createann.eq} obeys 	
\be a|n\ket^E=\sqrt n|n-1\ket^E, \ee we can derive the matrix elements 
  \be { }^\ontt\bra\vv_1|a|\vv_2\ket^\ontt&=&\frac 1{2\pi} \ex{-i\vv_1 }\,g(\vv_1-\vv_2)\ ,\labell{ainphispace.eq}\\
 \hbox{and}\qquad\quad
  { }^\ontt\bra\vv_1|\adag|\vv_2\ket^\ontt&=&\frac 1{2\pi} \,\ex{
  i\vv_2 }\ \,g(\vv_1-\vv_2)\ , \eel{adaginphi.eq}
and from this, using Eqs.~\eqn{createann.eq}, we find the matrix elements of the operators \(x\) and \(p\) of the original quantum harmonic oscillator, in terms of the basis states \(|\vv\ket^\ontt\):  
\be
  \bra\vv_1|x|\vv_2\ket&=&\frac 1{2\pi\sqrt 2}\big(\ex{-i\vv_1}+\ex{i\vv_2 } \big) \,g(\vv_1-\vv_2)\  ; \labell{xont.eq}\\\
 \bra\vv_1|p|\vv_2\ket&=&\frac i{2\pi\sqrt 2} \big( \ex{-i\vv_1}-\ex{i\vv_2} \big) \, g(\vv_1-\vv_2)\ . \eel{pont.eq}
 
 It  is possible to combine \(N\) oscillators with different frequencies  \(\w_i\)\,, requiring us to generalise equations \eqn{harmops.eq}  -- \eqn{harmeigen.eq} as
	\be H&=&\sum_{i=i}^N H_i\,; \qquad H_i=\half(\w_i^2 x_i^2+p_i^2-\w_i)\ \ =\  \w_i\adag_i a_i \ ,\labell{Homega.eq}\\
		a_i&=& \fract 1{\sqrt2}(\w_i^{1/2}\,x+i\w_i^{-1/2}\,p)\ ,\labell{aomega.eq}\\
		E_n^i&=&n_i\,\w_i\ .\eel{omega.eq}
This system of \(N\) quantum harmonic oscillators, gives us \(N\)  variables of the COV type,
	\be	\vv_i(t)&=&\vv_i(0)+\w_i t\mod 2\pi\ .\qquad\hbox{etc.} \eel{phit.eq}
	
Ideas of treating quantized field theories as systems in a box with periodic boundary conditions were investigated by Dolce\cite{Dolce.ref}. The wave equation then fixes the timelike component of the periodicities, and systems of this kind may then be regarded as multiple systems of COV variables. 

\newsecl{On the analytic structure of the auxiliary function \(G(z)\)}{auxiliary.sec}

The auxiliary function \(G(z)\) is defined by Eq.~\eqn{auxfn.eq}, but this only converges for values of \(z\) within the unit circle, that is, \(|z|<1\). Also, \emph{on} the unit circle, this definition seems to diverge. Usually, expansions that oscillate wildly at some distance from the origin, can be defined by slightly smearing the coefficients, but here, this procedure is tricky. Indeed, the mathematics needed to show that the probabilities generated by applying \(g(\vv)\) are uniquely defined and real, is rather delicate, an understatement, as shown in this section.

This section is intended only for mathematically minded readers. Their comments would be appreciated.

At finite \(N\) the function \be G_N(z)=\sum_{n=1}^N\sqrt n\, z^n\eel{GN.eq}
 has \(N\) zeros. Most of these will be close to the unit circle, \(|z|_n\ra 1\). The questions we would like to see answered are:

\bi{1.}  What will be the analytic structure of Eq.~\eqn{GN.eq} in the limit \(N\ra\infty\)?
\itm{2.} Is it possible at all to define and compute an analytic continuation for the function \(G_N\) for \(|z|>1\)?
\itm{3.} Where are the zeros and the poles of this analytic function?
\itm{4.} Can one prove that \be G^*(z)\qu G(z^*)\eel{G*z.eq} so that the operators \(x\) and \(p\) defined in Eqs~\eqn{xont.eq} and \eqn{pont.eq} can be seen to be hermitian?
\ei 
The last question is not quite trivial because one must first redefine the limit function \(g(\vv)\), but by careful study of the equations , we found that  Eq.~\eqn{G*z.eq} is true, due to the fact that the coefficients \(\sqrt n\) are all real, see Fig.~\ref{fg.fig}.

First, we find that \(G(z)\) is the second derivative of a function \(F(z)\) that stays strictly finite on the unit circle (where \(|z|=1\)):
\be G(z)=(z\pa_z)^2F(z)\ ,&& F(z)=\sum_{n=1}^\infty  z^n/(n\sqrt n)\,;\nm \\
g(\vv)=\frac{-\pa^2}{\pa\vv^2} \,f(\vv)\ ,&& f(\vv)=F(\ex{i\vv})\ . 
\eel{secder.eq}
 
\begin{figure}
\qqquad  \widthfig{260pt}{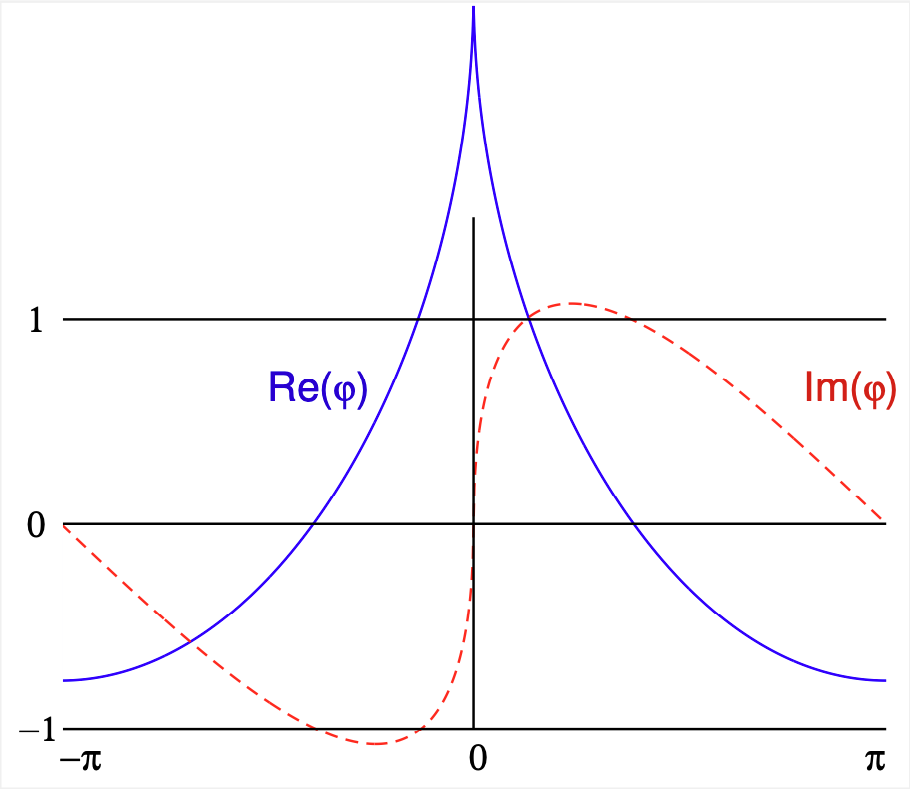}\caption{\small An accurate calculation of the function \(f(\vv)\), defined in Eq.~\ref{secder.eq}. Blue solid line: its real part, dashed red line: its imaginary part. Both real part and secondary part have a  divergence in their first and second derivatives, apparently only at the origin, \(\vv\ra 0\). Note that the imaginary part flips sign at \(\vv=\pm\pi\), by going through the branch cut. \labell{fg.fig} }\end{figure}

 \begin{figure} 
 \widthfig{450pt}{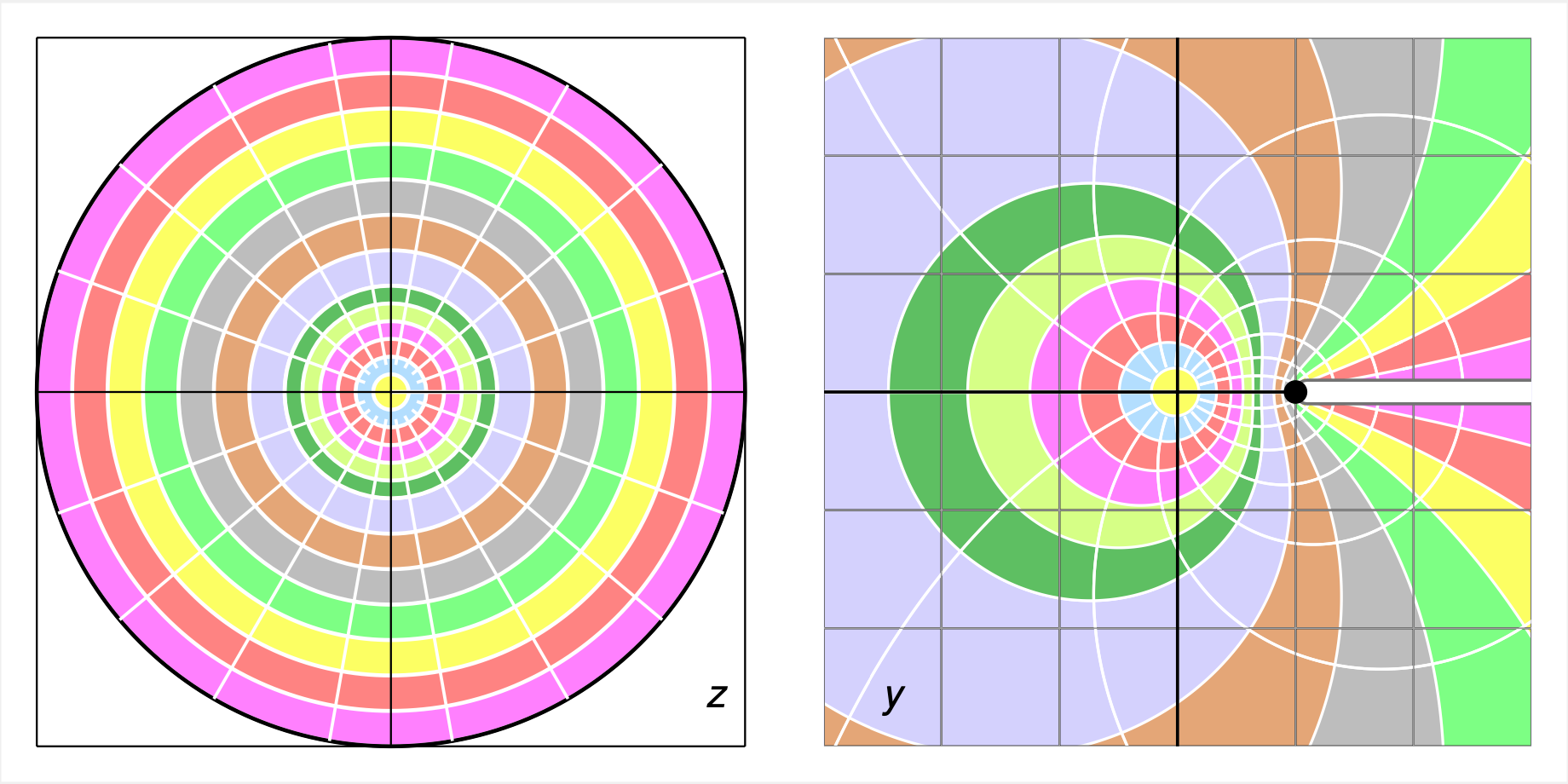}\caption{\small  a) \(z\) space, regions where the function \(G(z)\) converges: domains \(|z|\le .05,\ .1,\ \dots \  ,\ 1.0 \)  are shown. b) After the transformation \eqn{ytoz.eq}, these domains turn into the regions shown here, that is, the entire \(z\) plane up to the branch cut, will be singularity free if expressed in the new \(y\) variable.}\labell{sheets.fig}\end{figure}
 
Therefore, \(F(z)\) is also accurately defined \emph{on} the unit circle, but before using it to recuperate \(G(z)\), one must carefully choose the order of the limits \(N\ra\infty\) and \(|z|\ra 1\).   In practice, one encounters no problems, see Fig.~\ref{fg.fig}. 

One thing is very well visible: the imaginary part of the function \(f\) is \emph{anti}periodic if we rotate by an angle of \(2\pi\) in the \(z\)-plane. 
This means that there must be a branch cut if we try to extend the definition of \(f(\vv)\) to beyond the unit circle. The branch cut may be chosen to be a line from \(z=1\) to \(z=+\infty\) in the \(z\) plane. We choose the branch cut to start at \(z=1\),  because this is the most singular point of our function.
  	
	There will be periodicity over an angle of \(4\pi\), and therefore, the structure that we consider first, is a double Riemann sheet, the pairs of which are connected at the given branch cut.
	
	The correct double Riemann sheet  structure is obtained by the transformation
\be y=\frac {4z}{(1+z)^2}\ ,\eel{ytoz.eq}
and its inverse: \be z=-1+\frac 2y\big(1-\sqrt{1-y}\big)\ =\ \quart y +\fract 18 y^2+\ \cdots \ . \eel {ztoy.eq}
	This can also be written as 
\be \sqrt y =\frac 2{\sqrt z \ +\ 1/ {\sqrt z}}\ .\eel {symytoz.eq}
The second Riemann sheet describes the solution with the opposite sign of the square root on 
eq.~\eqn{ytoz.eq}.  There, we get the solution 
\be \tl z=-1+\frac 2y\big(1+\sqrt{1-y}\big)=1/z\ , \eel{ztilde}
	which is easiest to see in eq.~\eqn{symytoz.eq}.
	
Since Eq.~\eqn{GN.eq} holds on the first Riemann sheet, we now postulate the expansion 
\be G_N(z)=\sum_{n=1}^N\sqrt{n}\ \tl z^{\,n}=\sum_{n=1}^N\sqrt{n}\,z^{-n}\ ;\eel{2ndsheet.eq}	
	this is justified by noting that \(|z|<1\) in the first sheet, while \(|\tl z|=1/|z|<1\) on the second.

Finally, we must check whether the two sheets match on the unit circle, where \(|z|=1/|\tl z|=1\ .\) This is found to hold, by using \(1/z=z^*\) on the boundary. Since \(1/z\) and \(z^*\) obey the same series expansion, this must be true if the series converges on the boundary.	
	
	Fig.~\ref{sheets.fig} shows how the unit circle (Fig.~\ref{sheets.fig}a) is mapped on the first Riemann sheet (fig.~\ref{sheets.fig}b). by the function \eqn{ytoz.eq}, and how the branch cut at the right connects the two sheets. This answers questions (2) and (3): there is only a branch cut, no poles. The function \(G(z)\) does go to infinity where the branch cut begins; the function \(F\) stays finite. They are related through Eqs~\eqn{secder.eq}. Question (4) is now obviously answered in the positive.

\newsecl{Epilogue}{epi}

We showed how one may consider the quantum harmonic oscillator as an ontological theory in disguise. This is important since it appears to contradict theorems  claiming  that such a behaviour in quantum theories is impossible. Of course those theories were assumed to be far more general than a single harmonic oscillator, or even a simple collection of harmonic oscillators, but this now is a question of principle. Where is the dividing line? Which other quantum systems allow for the definition of COV variables, variables that commute with themselves and others at all times?  If for instance one considers the quantum field theory of bosonic free particles in a box of an arbitrary shape in multiple dimensions, one may observe that this is merely a collection of harmonic oscillators. 

One would be tempted to conclude that, therefore, bosonic particles in a box should also contain COV states\cite{GtHmassive.ref}, but there is a complication in such systems: it is not easy to restore locality in the COV, since they are defined in momentum space. Turning these into variables that are local in position space appears not to be impossible, but then there is another complication: the operators one obtains that way seem to violate Lorentz invariance. This happens since the box is not Lorentz invariant. It is conceivably possible to restore Lorentz invariance, but we presently do not know how to do this in the Standard Model. 

Thus our observations do not imply that text books on quantum mechanics have to be rewritten, except where they state explicitly that classical ontological variables cannot exist. Are \emph{local} ontological variables forbidden? Locality is a meaningless concept in a single quantum harmonic oscillator. In this paper we show exactly what an ontological variable \emph{is}. Emphatically, the ontological variable may be assumed to have a probability distribution as in quantum mechanics \emph{and} in classical theories: 

\begin{center} All uncertainties in the final state merely\\ reflect the uncertainties in the initial state. \end{center}

 As soon as we claim that the initial state is exactly given, the wave function of the final state will collapse. \emph{The harmonic oscillator requires no special axiom for the collapse of the wave function }-- provided that we stick to the observables in \(\vv\) space.
There, we do not need to assume the existence of many universes. Just one universe, ours, is all we need to understand.

We emphasise that what we found here as a modification of the usual picture of   quantum mechanics, is presumably merely the tip of an iceberg. It will not only apply to pure harmonic quantum oscillators, but also to many systems that evolve and interact in more generic ways. It is the fact that harmonic oscillators are periodic that counts. Whenever we consider a simplified model of nature where variables become periodic (for instance if we consider a box with periodic boundary conditions), one may observe that the energy spectrum consists of regular sequences of spectral lines (see Fig.~\ref{period.fig}), so that harmonically oscillating fields enter the picture. Time-periodic motion is always classical. All we then need to talk about then is how the probability distributions evolve.

In all classical systems, probability distributions evolve in the same orbits as the classical variables do. Consequently: \emph{probability in} = \emph{probability out}. If, in \(\vv\) space, the initial state is defined with infinite precision, the final state will also be infinitely precise. This implies that the `typically quantum feature' of the collapse of the wave function, has its counter part in ontological theories. In the model we presented, the variables \(\vv\) may be assumed to be infinitely sharply defined, but then also the final states will still be completely sharply defined; they always come in a collapsed form. 

The clash with usual findings concerning the `impossible' physical reality of quantum mechanical phenomena and calculations, lies in the fact that the duality transformation is only applicable in \emph{one} basis of Hilbert space: the one consisting of the ontological states. Choosing the conventional basis elements does not modify the results. The fact that we wish to emphasise is that, this `ontological' basis also never needs to be departed from, other than in approximative calculations: both the initial states and the final, observed states of any quantum process will be totally determined by the probabilities in the  genuinely ontological basis; therefore, other choices of basis will never  be necessary from a strictly logical viewpoint.

And it seems as if this possibility has never been considered before; however, see Refs. \cite{Brans-1987.ref} and 
\cite{Vervoort-2013.ref}. As for the numerous `quantum paradoxes' that have been formulated in the literature, the procedure needed, to formulate the probability patterns in an ontological basis, has been worked out in Ref.~\cite{GtHFF.ref}. The guiding principle: always stay in the ontological basis.

This brings me to the most  frequently uttered objections against the conclusions of this paper: \emph{We proved that the ontological theory mimics the quantum theory} only \emph{if we stay in the ontological basis}.
So what happens if I start in the \(x\) basis or in the \(p\) basis? Aren't there much more quantum states than classical states?!

The answer to this is no! An ontological theory such as the quantum harmonic oscillator, never collapses into states other than the eigenstates of an ontological operator.  If we start in an eigen state of a non ontological variable, we can always first calculate the probabilities for all ontological states, find that the series of probabilities obtained will not change in time, and conclude that the final state will consist of the same list of probabilities\fn{The quantum \emph{phases} then play no role anymore.}. The best experiments will always be the ones where one starts with one given ontological state, to discover that the final state will be collapsed into a single ontological state as well. Indeed, in contrast with any other approach to hidden variable theories, returning to the conventional quantum states will never give us more, or ``different'' states than the ontological ones that we already found.

The author benefitted from many discussions, notably with T. Palmer, C. Wetterich, M. Welling and D. Dolce.

 \end{document}